\documentclass[preprint,pre,aps,amsfonts,amsmath,superscriptaddress,nofootinbib,showkeys]{revtex4}
\usepackage{amsfonts,amsmath,amssymb,amscd}
\usepackage{mathrsfs}
\usepackage{empheq}
\usepackage{graphicx}
\usepackage{bm,bbold}
\usepackage{color}
\usepackage{calligra}
\usepackage[T1]{fontenc}
\usepackage{scalerel}
\newcommand*{\paral}{\stretchrel*{\parallel}{\perp}}
\begin{document}
%
%
\title{Testing the Instanton Approach to the Large Amplification Limit of a Diffraction-Amplification Problem}
\author{Philippe Mounaix}
\email{philippe.mounaix@polytechnique.edu}
\affiliation{CPHT, CNRS, \'Ecole
polytechnique, Institut Polytechnique de Paris, 91120 Palaiseau, France.}
\date{\today}
\begin{abstract}
The validity of the instanton analysis approach is tested numerically in the case of the diffraction-amplification problem $\partial_z\psi -\frac{i}{2m}\partial^2_{x^2} \psi =g\vert S\vert^2\, \psi$ for $\ln U\gg 1$, where $U=\vert\psi(0,L)\vert^2$. Here, $S(x,z)$ is a complex Gaussian random field, $z$ and $x$ respectively are the axial and transverse coordinates, with $0\le z\le L$, and both $m\ne 0$ and $g>0$ are real parameters. We consider a class of $S$, called the `one-max class', for which we devise a specific biased sampling procedure. As an application, $p(U)$, the probability distribution of $U$, is obtained down to values less than $10^{-2270}$ in the far right tail. We find that the agreement of our numerical results with the instanton analysis predictions in Mounaix (2023 {\it J. Phys. A: Math. Theor.} {\bf 56} 305001) is remarkable. Both the predicted algebraic tail of $p(U)$ and concentration of the realizations of $S$ onto the leading instanton are clearly confirmed, which validates the instanton analysis numerically in the large $\ln U$ limit for $S$ in the one-max class.
\end{abstract}
\maketitle
%
%
\section{Introduction}\label{intro}
In a recent work\ \cite{Mounaix2023}, the instanton analysis approach\ \cite{Raja1982,SS1998,FKLM1996} was used to determine the tail of $p(U)$, the probability distribution of $U=\vert\psi(0,L)\vert^2$, for $\ln U\gg 1$, $\psi(x,z)$ being the solution to the diffraction-amplification problem\footnote{This problem is of interest in, e.g., laser-plasma interaction and nonlinear optics in which $\psi$ is the complex time-envelope of the scattered light electric field, $g$ and $S$ being proportional to the average laser intensity and the complex time-envelope of the laser electric field, respectively\ \cite{RD1994}.}:
\begin{equation}\label{withDeq}
\left\lbrace
\begin{array}{l}
\partial_z\psi(x,z)-\frac{i}{2m}\partial^2_{x^2} \psi(x,z)=g\vert S(x,z)\vert^2\psi(x,z), \\
0\le z\le L,\ x\in\Lambda\subset\mathbb{R},\ {\rm and}\ \psi(x,0)=1.
\end{array}\right.
\end{equation}
Here, $z$ and $x$ respectively denote the axial and transverse coordinates in a domain of length $L$ and (one-dimensional) cross-section $\Lambda$. For technical convenience, we will take for $\Lambda$ the circle of length $\ell$. The boundary condition at $z=0$ is taken to be a constant for simplicity. Both $m\ne 0$ and $g>0$ are real parameters and $S(x,z)$ is a transversally homogeneous complex Gaussian noise with zero mean and normalization $L^{-1}\int_0^L \langle\vert S(x,z)\vert^2\rangle\, dz =1$.

From the instanton analysis of the corresponding Martin-Siggia-Rose action, it was found in\ \cite{Mounaix2023} that $S$ concentrates onto long filamentary instantons, $S_{\rm inst}$, as $\ln U\to +\infty$. These filamentary instantons run along specific non-random paths, denoted by $x_{\rm inst}(\cdot)$, that maximize the largest eigenvalue $\mu_1\lbrack x(\cdot)\rbrack$ of the covariance operator $T_{x(\cdot)}$ defined by
\begin{equation}\label{covariance3Dpath}
(T_{x(\cdot)} f)(z) =\int_0^L
C(x(z)-x(z^\prime) ,z,z^\prime)\, f(z^\prime)\, dz^\prime ,
\ \ \ f(z)\in L^2([0,L]),
\end{equation}
where $C(x-x^\prime ,z,z^\prime)=\langle S(x,z)S(x^\prime ,z^\prime)^\ast \rangle$. In equation~(\ref{covariance3Dpath}), $x(\cdot)$ is a continuous path in $\Lambda$ satisfying $x(L)=0$, and for the class of $S$ considered in\ \cite{Mounaix2023}, every maximizing path $x_{\rm inst}(\cdot)$ is also continuous with $x_{\rm inst}(L)=0$ (see\ \cite{Mounaix2023} for details). In the most common case of `single-filament instantons' for which there is only one maximizing path, and assuming a non-degenerate $\mu_1\lbrack x_{\rm inst}(\cdot)\rbrack$, one has $S(x,z)\sim S_{\rm inst}(x,z)$ $(\ln U\to +\infty)$ with
\begin{equation}\label{summaryintro1}
S_{\rm inst}(x,z)=\frac{c_1}{\mu_{\rm max}}
\, \int_0^L C(x-x_{\rm inst}(z^\prime),z,z^\prime)
\, \phi_1(z^\prime)\, dz^\prime ,
\end{equation}
where $\mu_{\rm max}$ is short for $\mu_1\lbrack x_{\rm inst}(\cdot)\rbrack$, $\phi_1$ is the fundamental eigenfunction of $T_{x_{\rm inst}(\cdot)}$, and $c_1$ is a complex Gaussian random variable with $\langle c_1\rangle =\langle c_1^2\rangle =0$ and $\langle\vert c_1\vert^2\rangle =\mu_{\rm max}$. The tail of $p(U)$ for $\ln U\gg 1$ can then be deduced from the statistics of $S_{\rm inst}$ as given in equation~(\ref{summaryintro1}). The result is a leading algebraic tail $\propto U^{-\zeta}$ with $\zeta =(1+1/2\mu_{\rm max} g)$, modulated by a slow varying amplitude (slower than algebraic)\ \cite{Mounaix2023}.

In the absence of a mathematically rigorous proof, the need for checking the validity of these analytical results numerically cannot be overlooked. To this end, it is essential to have a good sampling of the realizations of $S$ for which $\ln U\gg 1$. Unfortunately, such realizations are extremely rare events, far beyond the reach of any direct sampling with a reasonable sample size. For instance, for the same Gaussian field $S$ and parameters as in section $5$ of\ \cite{Mounaix2023} and in the simple diffraction-free limit, $m^{-1}=0$, in which $U$ can be computed exactly, it can be checked that $p(\ln U\ge 10^3)=O(10^{-100})$. It is thus clearly unrealistic to expect that the asymptotic analytical results can be tested by direct numerical simulations. To gain access to the asymptotic regime and check the validity of the instanton analysis we need a specific approach. One possible strategy is to bias the underlying distribution of $S$ towards the outcomes of interest. In the case of nonlinear equations with additive noise, this has been successfully achieved by means of the `importance sampling algorithm'\ \cite{HM1956} frequently used in rare event physics (see e.g.\ \cite{Hartmann2014,HDMRS2018,HMS2019} and references therein). For the diffraction-amplification problem~(\ref{withDeq}) with $S$ admitting a single and non-degenerate instanton, like in equation~(\ref{summaryintro1}) (see the appendix), it turns out that a different, somewhat simpler, method can be devised, based on the existence of a nonlinear fit to numerical data giving a highly accurate approximation of $U$ as a function of $S$, when $\ln U$ is large. It is then possible to determine the extreme upper tail of $p(U)$ and the corresponding realizations of $S$ from numerical simulations. Development of the biased sampling procedure and its application to check the validity of the instanton analysis in the case of problem~(\ref{withDeq}) is the subject of the present work.

Before entering the details of the calculations, it is useful to summarize our main results.

\begin{itemize}
\item Let $T_C$ denote the covariance operator defined by
\begin{equation}\label{TCoperator}
(T_C f)(x,z) =\int_0^L\int_{\Lambda}
C(x-x^\prime ,z,z^\prime)\, f(x^{\prime},z^\prime)\, dx^{\prime}dz^\prime ,
\ \ \ f(x,z)\in L^2(\Lambda\times [0,L]).
\end{equation}
Write $r_{\paral}$, and $r_{\perp}$ the $L^2$-norms of the components of $T_C^{-1/2}S$ parallel and perpendicular to any given direction in $L^2$ close enough to the one maximizing $U$ at fixed $\| T_C^{-1/2}S\|_2$, which is unique for the class of $S$ we consider (see Section~\ref{subsec3a}). By analyzing a large number of numerical solutions to equation~(\ref{withDeq}), we identify the existence of an implicit equation relating $U$, $r_{\paral}$, and $r_{\perp}$  when $\ln U$ is large and all the other quantities characterizing $T_C^{-1/2}S$ are fixed. The reason why $T_C^{-1/2}S$ appears rather than $S$ will be made clear at the end of section~\ref{modelanddefs} and above equation~(\ref{nonlinearfit}). More specifically, writing $r_{\paral}=\sqrt{\eta}\cos\theta$ and $r_{\perp}=\sqrt{\eta}\sin\theta$, with $0\le\theta\le\pi/2$ and $\eta>0$, we check that our numerical data satisfy
\begin{equation}\label{impliciteq}
\cos\theta -\left(\frac{\log_{10}U}{a\eta -b}\right)^{\alpha}=0,
\end{equation}
with very good accuracy for $1820\le\log_{10}U\le 2030$ and $1000\le\eta\le 1100$, where $a=1.86428$, $b=25.7163$, and $\alpha\sim 0.5$ is a random exponent depending on the quantities characterizing $T_C^{-1/2}S$ other than $\theta$ and $\eta$.
\item From this result, we derive the expression of the conditional probability distribution $p(U,\theta\vert\varSigma_{\rm orv})$ valid for $\log_{10}U\gg 1$, where $\varSigma_{\rm orv}$ stands for all the random variables characterizing $T_C^{-1/2}S$ other than $\theta$ and $\eta$ (`orv' is short for `other random variables'). For a given $U$ with $\log_{10}U\gg 1$, we then obtain (i) $p(U\vert\varSigma_{\rm orv})$ by integrating $p(U,\theta\vert\varSigma_{\rm orv})$ numerically over $0\le\theta\le\pi/2$, and (ii) $p(\theta\vert U,\varSigma_{\rm orv})$ as $p(U,\theta\vert\varSigma_{\rm orv})/p(U\vert\varSigma_{\rm orv})$.
\item We draw the realizations of $S$ given $U$ with $\log_{10}U\gg 1$ according to the procedure defined by: (a) draw $\varSigma_{\rm orv}$; (b) derive $p(\theta\vert U,\varSigma_{\rm orv})$; (c) draw $\theta$ from $p(\theta\vert U,\varSigma_{\rm orv})$ instead of from its unconditional probability distribution; and (d) use the resulting $T_C^{-1/2}S$ in $S=T_C^{1/2}(T_C^{-1/2}S)$ to get $S$.
\item We estimate $p(U)$ as the sample mean of $p(U\vert\varSigma_{\rm orv})$ over realizations of $\varSigma_{\rm orv}$ for different values of $U$ with $\log_{10}U\gg 1$ and we compare the results with the predictions of the instanton analysis. Let $\mathfrak{D}$ denote the $L^2$-distance between $S/\| S\|_{2}$ and $S_{\rm inst}/\| S_{\rm inst}\|_{2}$. For a given $U$ with $\log_{10}U\gg 1$, we estimate $p(\mathfrak{D}\vert U)$ as a sample mean over realizations of $\varSigma_{\rm orv}$ and $\theta$ (see equation~(\ref{pofdlessdelta}) and below). Finally, we plot $\vert S\vert^2/\| S\|^2_{2}$ for different realizations of $S$ with particular values of $\mathfrak{D}$ as well as the sample mean of $\vert S\vert^2/\| S\|^2_{2}$ over realizations of $S$ with close values of $\mathfrak{D}$, and we compare the results with the theoretical instanton profile.
\end{itemize}

The outline of the paper is as follows. The Gaussian field $S$ that we consider is specified in section\ \ref{modelanddefs}, where we also recall some results of\ \cite{Mounaix2023} needed in the sequel. Section\ \ref{biasprocedure} gives preliminary numerical results used in section\ \ref{application}. Finally, in section\ \ref{application} we define our biased sampling procedure and use it to test the instanton analysis of problem~(\ref{withDeq}) numerically, in the limit of a large $\ln U$ and for the class of $S$ considered.
%
%
\section{Model and definitions}\label{modelanddefs}
Since the present work is the continuation of the numerical study initiated in\ \cite{Mounaix2023}, section~5, we consider the same Gaussian field $S(x,z)$. Namely, we take
\begin{equation}\label{solSnum}
S(x,z)=\sum_{\mathclap{n=-50}}^{50}{\vphantom{\sum}}^\prime
s_{n}\sqrt{\varsigma_n}\, 
\exp\, i\left\lbrack\frac{2\pi n}{\ell} x+\left(\frac{2\pi n}{\ell}\right)^2\frac{z}{2}\right\rbrack ,
\end{equation}
where $\sum_n^{\, \prime}$ means $\sum_{n\ne 0}$. Here, the $s_n$s are independent and identically distributed (i.i.d.) complex Gaussian random variables with $\langle s_n\rangle =\langle s_n s_m\rangle =0$ and $\langle s_n s_m^\ast\rangle =\delta_{nm}$. The spectral density $\varsigma_n$ normalized to $\sum_{n=-50}^{\, \prime\, 50} \varsigma_n=1$ is given by the Gaussian spectrum
\begin{equation}\label{gaussspectrum}
\varsigma_{n}\propto\exp\left\lbrack -\left(\frac{\pi n}{\ell}\right)^2\right\rbrack .
\end{equation}
Equation~(\ref{solSnum}) is reminiscent of models of spatially smoothed laser beams\ \cite{RD1993}, where $S$ is a solution to the paraxial wave equation
\begin{equation}\label{eqSnum}
\partial_z S(x,z)+\frac{i}{2}\partial^2_{x^2} S(x,z)=0,
\end{equation}
with boundary condition $S(x,0)=\sum_{n=-50}^{\, \prime\, 50} s_n\sqrt{\varsigma_n}\, \exp(2i\pi nx/\ell)$. The $n=0$ mode is excluded from the Fourier representation~(\ref{solSnum}) to ensure that the space average $\ell^{-1}\int_{\Lambda}S(x,z)\, dx$ is zero for all $z$ and every realization of $S$, as expected for the electric field of a smoothed laser beam.

For each realization of $S$ on a cylinder of length $L=10$ and circumference $\ell =20$, we solve equation~(\ref{withDeq}) by using a symmetrized $z$-split method \cite{Strang1968} which propagates the diffraction term, $(i/2m)\partial^2_{x^2}\psi(x,z)$, in Fourier space and the amplification term, $g\vert S(x,z)\vert^2\psi(x,z)$, in real space. We take $m=0.7$, and $g=0.5$. For $S$ in equation~(\ref{solSnum}) with Gaussian spectrum~(\ref{gaussspectrum}) and the given values of $L$ and $\ell$, it is shown in\ \cite{Mounaix2023} that:
\medskip

(i) there is a single instanton, $S_{\rm inst}$, which is a single-filament instanton of the form\ (\ref{summaryintro1}) in which $x_{\rm inst}(\cdot)\equiv 0$, $\mu_{\rm max}=4.34984$, and
\begin{equation}\label{corfunction}
C(x,z,z^\prime)=C(x,z-z^\prime)=
\sum_{\mathclap{n=-50}}^{50}{\vphantom{\sum}}^\prime \varsigma_n\, 
\exp\, i\left\lbrack\frac{2\pi n}{\ell} x+\left(\frac{2\pi n}{\ell}\right)^2\frac{z-z^\prime}{2}\right\rbrack ;
\end{equation}
\medskip

(ii) the convolution representation of $S_{\rm inst}$ in equation~(\ref{summaryintro1}) with $x_{\rm inst}(\cdot)\equiv 0$ and $C$ in equation~(\ref{corfunction}) is equivalent to the Fourier representation
\begin{equation}\label{instsolSnum}
S_{\rm inst}(x,z)=\sum_{\mathclap{n=-50}}^{50}{\vphantom{\sum}}^\prime
\mathfrak{s}_{n}\sqrt{\varsigma_n}\, 
\exp\, i\left\lbrack\frac{2\pi n}{\ell} x+\left(\frac{2\pi n}{\ell}\right)^2\frac{z}{2}\right\rbrack ,
\end{equation}
where $\bm{\mathfrak{s}}$ (with components $\mathfrak{s}_{n}$) is an eigenvector of the $100\times 100$ positive definite Hermitian matrix
\begin{equation}\label{matrixM}
M_{nm}=\sqrt{\varsigma_n \varsigma_m}\, \int_0^L \exp i\left\lbrack\left(\frac{2\pi}{\ell}\right)^2
(m^2-n^2)\, \frac{z}{2}\right\rbrack\, dz,\ \ \ \ (n,m\ne 0),
\end{equation}
associated with the eigenvalue $\mu_{\rm max}$. (Note that $M$ and $T_{x_{\rm inst}(\cdot)\equiv 0}$ have the same eigenvalues with the same degeneracies\ \cite{MCL2006}). The $\mathfrak{s}_{n}$s in equation~(\ref{instsolSnum}) are correlated complex Gaussian random variables with $\langle \mathfrak{s}_{n}\rangle =\langle \mathfrak{s}_{n}\mathfrak{s}_{m}\rangle =0$ and $\langle \mathfrak{s}_{n}\mathfrak{s}_{m}^\ast\rangle =\mathfrak{e}_{n}^{(1)}\mathfrak{e}^{(1)\, \ast}_{m}$, where $\bm{\mathfrak{e}}^{(1)}$ (with components $\mathfrak{e}_{n}^{(1)}$) is the normalized fundamental eigenvector of $M$ (see\ \cite{Mounaix2023}, section 4.1, for details).
\medskip

\begin{figure}[!h]
\includegraphics[width = 0.9\linewidth]{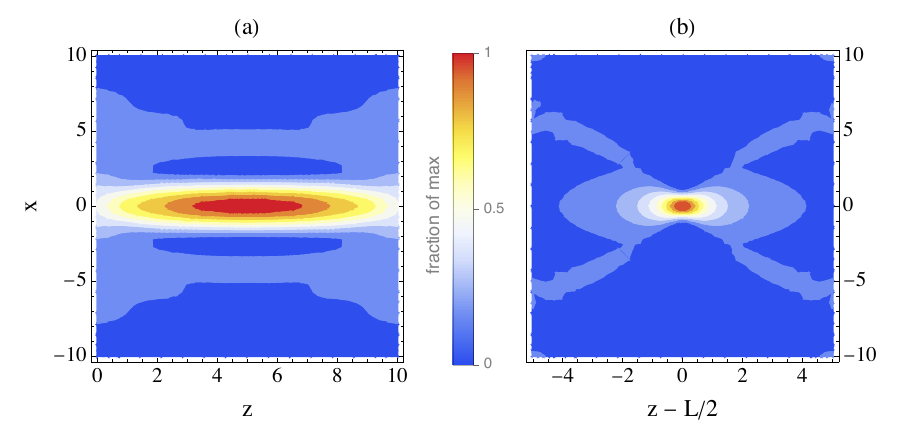}
\caption{(a) Contour plot of $\vert S_{\rm inst}(x,z)\vert^2$ for $S_{\rm inst}(x,z)$ in equation~(\ref{instsolSnum}). (b) Contour plot of the autocorrelation profile (or `hot spot profile') $\vert C(x,z-L/2)\vert^2$ for $C(x,z-z^\prime)$ in equation~(\ref{corfunction}). (Color scale legend : $1$ corresponds to the maximum of the plotted function.)}\label{figureinst}
\end{figure}

Figure~\ref{figureinst} shows the contour plots of $\vert S_{\rm inst}\vert^2$ and the autocorrelation profile $\vert C\vert^2$ (also referred to as `hot spot profile' in the laser-matter interaction literature\ \cite{RD1993}).

Define $\hat{S}_{\rm inst}= S_{\rm inst}/\| S_{\rm inst}\|_{2}$ and $\hat{S}=S/\| S\|_{2}$ where $\|\cdot\|_{2}$ denotes the $L^2$-norm on $\Lambda\times\lbrack 0,L\rbrack$. Write $S_{\parallel}=\left(\hat{S}_{\rm inst},\hat{S}\right) \hat{S}_{\rm inst}$ the component of $\hat{S}$ along $S_{\rm inst}$. We measure the departure of the realizations of $S$ from the instanton through the $L^2$-distance
\begin{equation}\label{distnum1}
\mathfrak{D}=\|\hat{S}-S_{\parallel}\|_{2}
=\sqrt{1-\vert(\hat{S},\hat{S}_{\rm inst})\vert^2}.
\end{equation}
Using the Fourier representations (\ref{solSnum}) and (\ref{instsolSnum}) in which we write $\bm{s}$ (with components $s_{n}$) as $\bm{s}=\|\bm{s}\|\, \hat{\bm{s}}$ and $\bm{\mathfrak{s}}=\|\bm{\mathfrak{s}}\|\, \hat{\bm{\mathfrak{s}}}$ with $\hat{\bm{\mathfrak{s}}}={\rm e}^{i\, {\rm arg}(c_1)}\bm{\mathfrak{e}}^{(1)}$, where $\|\cdot\|$ is the usual Euclidean norm and where we have used $\bm{\mathfrak{s}}=\mu_{\rm max}^{-1/2} c_1 \bm{\mathfrak{e}}^{(1)}$ (see the first equation~(52) in\ \cite{Mounaix2023}), one gets
\begin{equation}\label{distnum2}
\mathfrak{D}=\sqrt{1-
\frac{\vert\sum_{n}^{\, \prime}\varsigma_n \hat{s}_n\mathfrak{e}_n^{(1)\, \ast}
\vert^2}{\left(\sum_{n}^{\, \prime}\varsigma_n \vert\hat{s}_n\vert^2\right)\, 
\left(\sum_{n}^{\, \prime}\varsigma_n \vert \mathfrak{e}_n^{(1)}\vert^2\right)}},
\end{equation}
which is the counterpart of the equation (66) in\ \cite{Mounaix2023} for fixed $y=0\, $\footnote{Note the typo on the right-hand side of equation (66) in\ \cite{Mounaix2023}: in the denominator, it should read $\hat{s}_n$ and $\hat{\mathfrak{s}}_n={\rm e}^{i\, {\rm arg}(c_1)}\mathfrak{e}_n^{(1)}$ instead of $s_n$ and $\mathfrak{s}_n$.}.

From equations~(\ref{TCoperator}), (\ref{solSnum}), and (\ref{corfunction}) it can be checked that the $s_n$s are also the components of $T_C^{-1/2}S$ in the orthonormal function basis $(1/\sqrt{\ell L})\, \exp i\lbrack 2\pi nx/\ell +(2\pi n/\ell)^2 z/2\rbrack$, which trivially defines an isomorphism between the $T_C^{-1/2}S$-space with $L^2$-inner product and the $\bm{s}$-space with usual dot product. As we will see shortly, for $S$ given by equation~(\ref{solSnum}) our results come out naturally in terms of $\bm{s}$, and therefore $T_C^{-1/2}S$, which explains why it is $T_C^{-1/2}S$ that appears in the summary at the end of section~\ref{intro}, rather than $S$.

We now have everything we need to move on to the presentation of our biased sampling method and its application to check the validity of the results obtained in\ \cite{Mounaix2023}. This is the subject of the next two sections.
%
%
\section{Preliminary results}\label{biasprocedure}
The problem is to sample large values of $\log_{10}U$. To this end, we need a biased sampling of the realizations of $S$ or, equivalently, of the random vector ${\bm s}$. The relationship between $U$ and ${\bm s}$ is extremely intricate and no simple general expression of the form $U=U({\bm s})$ can be written explicitly. All we can say at this stage is that a large $\log_{10}U$ implies a large $\|\bm{s}\|$, while the converse is not true\footnote{A realization of $\bm{s}$ with a large $\|\bm{s}\|$ can yield a large $\log_{10}\vert\psi(x,L)\vert^2\gg 1$ at some $x$ away from zero and $\log_{10}U=\log_{10}\vert\psi(0,L)\vert^2=O(1)$ even though $\|\bm{s}\|$ is large.}. So, sampling large $\|\bm{s}\|$'s alone is not sufficient.

According to instanton analysis, the larger $\log_{10}U$ the more $\bm{s}$ tends to align with $\bm{\mathfrak{e}}^{(1)}$. Testing the instanton analysis in the large $\log_{10}U$ regime therefore requires as a prerequisite to be able to control the direction of $\bm{s}$, which can be achieved by a change of variables making the direction and amplitude of $\bm{s}$ explicit.

Write $N$ the number of terms in the sum on the right-hand side of equation~(\ref{solSnum}) ($N=100$). A given realization of $S$ corresponds to a given realization of the (complex) $N$-dimensional vector $\bm{s}$ with coordinates $s_{n}$, and conversely. Let $\hat{\bm{u}}$ be a given $N$-dimensional (complex) unit vector, not necessarily equal to $\bm{\mathfrak{e}}^{(1)}$. Define $\bm{s}_{\paral}=(\bm{s}\cdot\hat{\bm{u}}^\ast)\, \hat{\bm{u}}=r_{\paral}{\rm e}^{i\varphi}\hat{\bm{u}}$ and $\bm{s}_{\perp}=\bm{s}-\bm{s}_{\paral}=r_{\perp}\, \hat{\bm{s}}_{\perp}$ with $r_{\perp} =\|\bm{s}_{\perp}\|$. Switching to polar coordinates $r_{\paral}=\sqrt{\eta}\cos\theta$ and $r_{\perp}=\sqrt{\eta}\sin\theta$, with $\eta =\|\bm{s}\|^2$ and $0\le\theta\le\pi/2$, we characterize the realizations of $S$ by the new variables $\theta$, $\eta$, $\varphi$, and $\hat{\bm{s}}_{\perp}$. The polar angle $\theta$ measures how close the direction of $\bm{s}$ is to that of $\hat{\bm{u}}$: $\theta=0$ means that $\bm{s}$ is aligned with $\hat{\bm{u}}$ ($\bm{s}_{\perp}=0$), whereas $\theta=\pi/2$ means that $\bm{s}$ is orthogonal to $\hat{\bm{u}}$ ($\bm{s}_{\paral}=0$).

The statistical properties of the new variables are deduced from the ones of the $s_n$s. Since the $s_{n}$s are i.i.d. standard complex Gaussian random variables, the projection of $\bm{s}$ onto any given direction is also a standard complex Gaussian random variable independent of the projections onto the orthogonal directions. This applies in particular to $s_{\paral}=(\bm{s}\cdot\hat{\bm{u}}^\ast)$ and the components of $\bm{s}_{\perp}$. After some straightforward algebra, one finds that the probability distribution functions (pdf) of $\theta$ and $\eta$ are respectively given by
\begin{equation}\label{pdftheta}
f(\theta)=2(N-1)\, (\sin\theta)^{2N-3}\cos\theta\, \bm{1}_{0\le\theta\le\pi/2} ,
\end{equation}
and
\begin{equation}\label{pdfeta}
h(\eta)=\Gamma(N)^{-1}\, \eta^{N-1} {\rm e}^{-\eta}\, \bm{1}_{0\le\eta}.
\end{equation}
The random phase $\varphi$ and tip of $\hat{\bm{s}}_{\perp}$ (i.e. the direction of $\bm{s}_{\perp}$)  are uniformly distributed over $\lbrack 0,2\pi)$ and the sphere $\|\bm{s}_{\perp}\| =1$, respectively.
%
%
\subsection{Closeness rule}\label{subsec3a}
The class of $S$ we consider is defined by the condition that, for fixed $\eta\gg 1$, there is only one direction of $\bm{s}$ that globally maximizes $\log_{10}U(\eta ,\hat{\bm{s}})$. Call this direction `the maximizing direction' and the class of $S$ `the one-max class'. It is shown in the appendix that $S$ as given by the equations\ (\ref{solSnum}) and\ (\ref{gaussspectrum}) and, more generally, any $S$ admitting a single and non-degenerate instanton,  belongs to this class. For a given $\hat{\bm{u}}$, there are two possibilities: either $\hat{\bm{u}}$ is the maximizing direction or it is not. The way we distinguish between these two possibilities is based on the following reasoning. If $\hat{\bm{u}}$ is the maximizing direction, then all the directions other than $\hat{\bm{u}}$ contribute negligibly to the amplification compared with $\hat{\bm{u}}$ (in the large  $\eta$ limit we consider), and the computed $\log_{10}U$ depends negligibly on the perpendicular components $\hat{\bm{s}}_{\perp}$. In addition, at $\theta =0$, $\bm{s}$ is aligned with $\hat{\bm{u}}$ ($\bm{s}_\perp =0$) and $\log_{10}U$ reaches its maximum, i.e., $\log_{10}U_{\theta =0}=\max_{\theta}\log_{10}U_{\theta}$ for all realizations of $\hat{\bm{s}}_\perp$. By contraposition, if the computed $\log_{10}U$ is found to be sensitive to $\hat{\bm{s}}_\perp$ or if there is a realization of $\hat{\bm{s}}_\perp$ for which $\log_{10}U_{\theta =0}<\max_{\theta}\log_{10}U_{\theta}$, then $\hat{\bm{u}}$ is not the maximizing direction. Conversely, if $\hat{\bm{u}}$ is not the maximizing direction, then $\log_{10}U$ does not reach its maximum at $\theta =0$ and there are realizations of $\hat{\bm{s}}_\perp$ for which $\log_{10}U_{\theta =0}<\max_{\theta}\log_{10}U_{\theta}$. In addition, for $\theta >0$, there are realizations of $\hat{\bm{s}}_\perp$ with a non negligible component along the maximizing direction, which makes the computed $\log_{10}U$ sensitive to $\hat{\bm{s}}_\perp$. Observing one of these two characteristics in the behavior of $\log_{10}U$ is sufficient to conclude that $\hat{\bm{u}}$ is not the maximizing direction. By contraposition, if the computed $\log_{10}U$ is observed to depend negligibly on $\hat{\bm{s}}_\perp$ and if $\log_{10}U_{\theta =0}=\max_{\theta}\log_{10}U_{\theta}$ for each realization of $\hat{\bm{s}}_\perp$, then $\hat{\bm{u}}$ is the maximizing direction.

For our purposes, we do not need to know whether $\hat{\bm{u}}$ is exactly the maximizing direction, but rather whether it is close enough to the maximizing direction, so that a good sampling of $\hat{\bm{s}}$ around $\hat{\bm{u}}$ ensures a good sampling of $\hat{\bm{s}}$ around the nearby maximizing direction, thereby providing access to large values of $\log_{10}U$ (for $\eta\gg 1$). This does not change the reasoning above which leads to the following `closeness rule' that we will use for deciding whether $\hat{\bm{u}}$ is close to the maximizing direction.

\bigskip
\noindent{\it Closeness rule}: for $\eta\gg 1$, $\hat{\bm{u}}$ is close to the maximizing direction {\it iff} $\log_{10}U$ for $\hat{\bm{s}}$ around $\hat{\bm{u}}$ is observed to depend very little on $\hat{\bm{s}}_\perp$ and $\log_{10}U_{\theta =0}=\max_{\theta}\log_{10}U_{\theta}$ for all the sampled realizations of $\hat{\bm{s}}_\perp$ to within a negligible fraction of them.

\bigskip
The point is that by applying this rule to numerical outcomes, we can check whether $\hat{\bm{u}}$ is close to the maximizing direction numerically without having to compare the values of $\log_{10}U$ for different $\hat{\bm{u}}$'s. It is worth noticing that the closeness rule applies to any $S$ in the one-max class, whatever the maximizing direction. It is not necessary to know it beforehand. For this class of $S$, the closeness rule does not presuppose any knowledge of the instanton solution to be tested.
%
%
\subsection{Setting a reference direction $\bm{\hat{u}}$}\label{subsec3b}
\begin{figure}[!h]
\includegraphics[width = 0.9\linewidth]{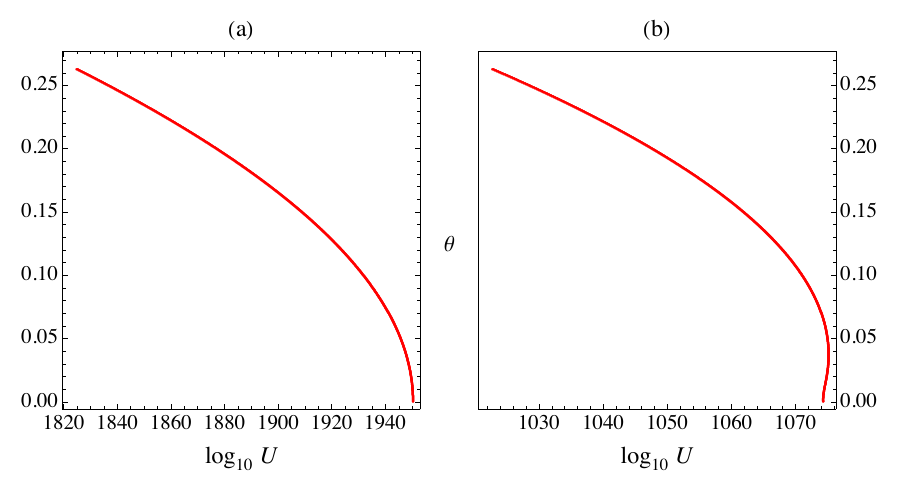}
\caption{Scatter plots of $\theta$ and $\log_{10}U$ for a given realization of $\varphi$ and $\hat{\bm{s}}_{\perp}$, fixed $\eta=1060$, and two different (orthogonal) reference directions: (a) $\hat{\bm{u}}=\bm{\mathfrak{e}}^{(1)}$ and (b) $\hat{\bm{u}}=\bm{\mathfrak{e}}^{(2)}$.}
\label{figurethetaofu1}
\end{figure}

\begin{figure}[!h]
\includegraphics[width = 0.6\linewidth]{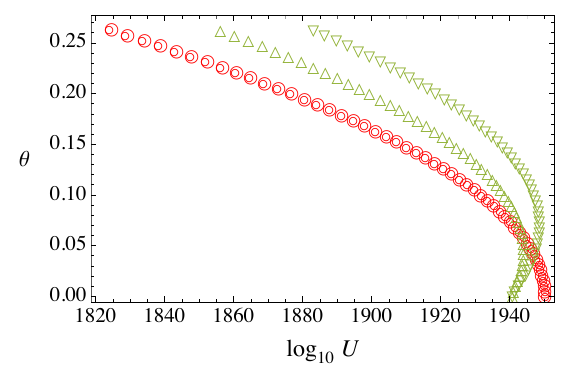}
\caption{Scatter plots of $\theta$ and $\log_{10}U$ for two different realizations of $\varphi$ and $\hat{\bm{s}}_{\perp}$, each represented by a specific marker. Large and small red circles are for $\hat{\bm{u}}=\bm{\mathfrak{e}}^{(1)}$ with $\eta=1060$, up and down green triangles correspond to $\hat{\bm{u}}=\bm{\mathfrak{e}}^{(2)}$ with $\eta=1880$. (For better legibility, only $50$ of the $10^3$ points actually computed are shown.)}
\label{figurethetaofu2}
\end{figure}

It is instructive to compare the values of $\log_{10}U$ and the compliance with the closeness rule for various reference directions $\hat{\bm{u}}$. We have made this comparison for $\hat{\bm{u}}=\bm{\mathfrak{e}}^{(1)}$ and the orthogonal directions $\hat{\bm{u}}=\bm{\mathfrak{e}}^{(q\ge 2)}$, where $\bm{\mathfrak{e}}^{(q)}$ is the normalized eigenvector of $M$ associated with the $q$th largest eigenvalue $\mu_q$ ($1\le q\le N$). (For the Gaussian spectrum~(\ref{gaussspectrum}), we have checked numerically that none of the $\mu_q$s is degenerate.) Figure\ \ref{figurethetaofu1} shows scatter plots of $\theta$ and numerically computed $\log_{10}U$ for a given realization of $\varphi$ and $\hat{\bm{s}}_{\perp}$, fixed $\eta=1060$, and $10^3$ ascending values of $\theta$ regularly spaced by $\Delta\theta=0.263\ 10^{-3}$ starting from $\theta=0$. Figures~\ref{figurethetaofu1}(a) and (b) correspond to $\hat{\bm{u}}=\bm{\mathfrak{e}}^{(1)}$ and $\hat{\bm{u}}=\bm{\mathfrak{e}}^{(2)}$, respectively (with $\mu_1=\mu_{\rm max}=4.34984$ and $\mu_2=2.0126$). It is clear that for a given $\eta\gg 1$, $\log_{10}U$ is significantly smaller in the case $\hat{\bm{u}}=\bm{\mathfrak{e}}^{(2)}$ (figure\ \ref{figurethetaofu1}(b)) than for $\hat{\bm{u}}=\bm{\mathfrak{e}}^{(1)}$ (figure\ \ref{figurethetaofu1}(a)). To have comparable values of $\log_{10}U$ we need a larger $\eta$ for $\hat{\bm{u}}=\bm{\mathfrak{e}}^{(2)}$ than for $\hat{\bm{u}}=\bm{\mathfrak{e}}^{(1)}$. To be more precise, for $\hat{\bm{u}}=\bm{\mathfrak{e}}^{(2)}$ we have checked that data with $\log_{10}U\gtrsim 1820$ --- the smallest value in figure\ \ref{figurethetaofu1}(a) --- require $\eta\gtrsim 1760$. For instance, it can be seen in figure~\ref{figurethetaofu2} that it takes $\eta\simeq 1880$ to bring the data for $\hat{\bm{u}}=\bm{\mathfrak{e}}^{(2)}$ in the same range of $\log_{10}U$ as for $\hat{\bm{u}}=\bm{\mathfrak{e}}^{(1)}$ in figure\ \ref{figurethetaofu1}(a). Given the fast decreasing tail of $h(\eta)$ in equation~(\ref{pdfeta}), an important consequence of this result is that among data falling in the same $(\log_{10}U,\theta)$ region as in figure\ \ref{figurethetaofu1}(a), the proportion of realizations of $\hat{\bm{s}}$ biased towards $\bm{\mathfrak{e}}^{(2)}$ is completely negligible compared to that of realizations biased towards $\bm{\mathfrak{e}}^{(1)}$, typically by a factor less than $h(1760)/h(1060)= O(10^{-282})$. We have checked that the proportion of realizations of $\hat{\bm{s}}$ biased towards $\bm{\mathfrak{e}}^{(q\ge 3)}$ is by far even smaller. From these first results, we can already conclude that the maximizing direction is not orthogonal to $\bm{\mathfrak{e}}^{(1)}$. To go further we will use the closeness rule.

Figure~\ref{figurethetaofu2} shows the same kind of scatter plots as in the figure~\ref{figurethetaofu1} for $\hat{\bm{u}}=\bm{\mathfrak{e}}^{(1)}$ with $\eta=1060$ and $\hat{\bm{u}}=\bm{\mathfrak{e}}^{(2)}$ with larger $\eta=1880$, each for two different realizations of $\varphi$ and $\hat{\bm{s}}_{\perp}$. For better legibility, only $50$ of the $10^3$ points actually computed are shown. Both figures\ \ref{figurethetaofu1} and\ \ref{figurethetaofu2} show that for a given realization of $\varphi$ and $\hat{\bm{s}}_{\perp}$, the data collapse on a single well-defined curve in the $(\log_{10}U,\theta)$ plane. For $\hat{\bm{u}}=\bm{\mathfrak{e}}^{(2)}$, it is clear that (i) $\log_{10}U_{\theta =0}<\max_{\theta}\log_{10}U_{\theta}$ and (ii) a change in the realization of $\hat{\bm{s}}_{\perp}$ visibly affects the curve (see up and down green triangles in figure\ \ref{figurethetaofu2}), which means that the contribution of $\bm{s}_{\perp}$ to the amplification is not negligible in this case. We have observed the same behavior for $\hat{\bm{u}}=\bm{\mathfrak{e}}^{(q\ge 3)}$ with an even more pronounced sensitivity to $\hat{\bm{s}}_{\perp}$. Furthermore, as $q$ increases, a dispersion of the data around the curve in the $(\log_{10}U,\theta)$ plane becomes visible and increases with $q$. These results do not meet the closeness rule, which implies that none of the orthogonal directions $\bm{\mathfrak{e}}^{(q\ge 2)}$ is close to the maximizing direction, in agreement with the discussion of figure~\ref{figurethetaofu1}.

In contrast, for $\hat{\bm{u}}=\bm{\mathfrak{e}}^{(1)}$ we observe that (i) $\log_{10}U_{\theta =0}$ is indistinguishable from $\max_{\theta}\log_{10}U_{\theta}$ (see also figures~\ref{figurethetaofu1} and\ \ref{figurethetaofu3}) and (ii) the curve depends very little on $\hat{\bm{s}}_{\perp}$ (small and large red circles in figure\ \ref{figurethetaofu2}). We have checked that the same was always true for several different subsamples of realizations of $\hat{\bm{s}}_{\perp}$. It can then be concluded from the closeness rule that $\bm{\mathfrak{e}}^{(1)}$ is close to the maximizing direction, so that sampling large $\eta$'s and $\hat{\bm{s}}$ around $\bm{\mathfrak{e}}^{(1)}$ simultaneously will yield a good sampling of the large values of $\log_{10}U$. Thus, it is legitimate to take $\hat{\bm{u}}=\bm{\mathfrak{e}}^{(1)}$ as a reference direction, which will be the case from now on.

The fact that the numerical results for $\hat{\bm{u}}=\bm{\mathfrak{e}}^{(1)}$ meet the closeness rule and validate $\bm{\mathfrak{e}}^{(1)}$ as being close to the maximizing direction is a necessary but not sufficient condition for the validity of the instanton analysis. It remains to find out how to use this result to achieve a good sampling of the large values of $\log_{10}U$ from which numerical results and analytical predictions can be compared. This is the subject of the sections\ \ref{subsec3c} and\ \ref{application}.

{\it Notational remark}: it may be useful to briefly come back to the notations used in sections~\ref{intro} and \ref{biasprocedure}. As explained at the end of section~\ref{modelanddefs}, $\bm{s}$ and $\bm{\mathfrak{s}}$ are respectively isomorphic to $T_C^{-1/2}S$ and $T_C^{-1/2}S_{\rm inst}$. Consequently, $r_{\paral}$ and $r_{\perp}$ are also the $L^2$-norms of the components of $T_C^{-1/2}S$ parallel and perpendicular to $T_C^{-1/2}S_{\rm inst}$ (in the $L^2$ sense) which is close to the maximizing direction, as written in section~\ref{intro}. The quantities characterizing $T_C^{-1/2}S$ other than $\theta$ and $\eta$, denoted by $\varSigma_{\rm orv}$ in section~\ref{intro}, are the random variable $\varphi$ and vector $\hat{\bm{s}}_{\perp}$.
%
%
\subsection{Highly accurate approximation of $\bm{\ln U(s)\gg 1}$ and applications}\label{subsec3c}
\begin{figure}[!h]
\includegraphics[width = 0.6\linewidth]{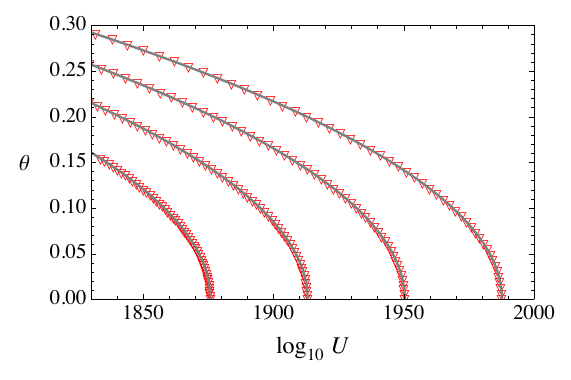}
\caption{Scatter plots of $\theta$ and $\log_{10}U$ for a given realization of $\varphi$ and $\hat{\bm{s}}_{\perp}$, and four different values of $\eta$ (down red triangles). Solid lines are plots of the nonlinear fit~(\ref{nonlinearfit}) for the corresponding $\eta$ and numerically computed $\alpha(\bm{s})$. Parameter values are $(\eta, \alpha(\bm{s}))=(1020,0.52569)$, $(1040,0.524924)$, $(1060,0.524862)$, and $(1080,0.525024)$, from left to right. (In each plot, only $50$ of the $10^3$ points actually computed are shown.)}
\label{figurethetaofu3}
\end{figure}

In figure\ \ref{figurethetaofu3} we show scatter plots of $\theta$ and $\log_{10}U$  for a given realization of $\varphi$ and $\hat{\bm{s}}_{\perp}$, and four different values of $\eta$ (see caption for details). Solid lines correspond to the nonlinear fit
\begin{equation}\label{nonlinearfit}
\theta =\cos^{-1}\left(\frac{\log_{10}U}{a\eta -b}\right)^{\alpha(\bm{s})},
\end{equation}
where $a=1.86428$, $b=25.7163$, and $\alpha(\bm{s})$ is a realization dependent exponent. Data points and nonlinear fits match remarkably well. We have checked on $10^2$ realizations of $\varphi$ and $\hat{\bm{s}}_{\perp}$, $10$ values of $\eta$ between $1010$ and $1100$, and $10^3$ values of $\theta$ like in figure\ \ref{figurethetaofu1} (which represents a total of $10^6$ different realizations of $S$), that for each $\varphi$, $\hat{\bm{s}}_{\perp}$, and $\eta$, the $10^3$ data points and the nonlinear fit are practically indistinguishable over the whole range $1820\le\log_{10}U\le 2030$. Numerical results also show that (i) no systematic (monotonic) variation of $\alpha(\bm{s})$ with $\eta$ at fixed $\varphi$ and $\hat{\bm{s}}_{\perp}$ is observed in the range of $\eta$ considered, and (ii) the relative dispersion of $\alpha(\bm{s})$ for different values of $\eta$ at fixed $\varphi$ and $\hat{\bm{s}}_{\perp}$ is one order of magnitude less than for different realizations of $\varphi$ and $\hat{\bm{s}}_{\perp}$ at fixed $\eta$ : $\Delta\alpha(\bm{s})/\langle\alpha(\bm{s})\rangle =10^{-3}$ and $10^{-2}$, respectively. Thus, with a good accuracy level, it is not unreasonable to ignore the dependence of $\alpha(\bm{s})$ on $\eta$ and write $\alpha(\bm{s})=\alpha(\varphi ,\hat{\bm{s}}_{\perp})$. We have checked by replacing $\alpha(\bm{s})$ in equation~(\ref{nonlinearfit}) with values obtained for different $\eta$ and the same realization of $\varphi$ and $\hat{\bm{s}}_{\perp}$, all other things being equal, that the error made in the nonlinear fit is indeed imperceptible.

We now use these results to write $p(U)$ and conditional probability $P(\mathcal{A}\vert U)$, where $\mathcal{A}$ is a set of realizations of $\bm{s}$, in forms suitable to numerical estimates in the asymptotic regime $\log_{10}U\gg 1$. Write $U(\bm{s})=\vert\psi_{\bm{s}}(0,L)\vert^2$, $\psi_{\bm{s}}(x,z)$ being the solution to equation~(\ref{withDeq}) for a given $S$ (hence $\bm{s}$). Our numerical results show that inverting equation~(\ref{nonlinearfit}) gives a highly accurate approximation of $U(\bm{s})$ when $\log_{10}U\gg 1$. Replacing $U(\bm{s})$ with this approximation in the exact expression $p(U)=\left\langle\delta(U-U(\bm{s}))\right\rangle_{\bm{s}}$, where $\langle\cdot\rangle_{\bm{s}}$ denotes the average over the realizations of $\bm{s}$, and integrating out $\eta$, one obtains
\begin{eqnarray}\label{pofuintegral}
p(U)&=&\frac{1}{\ln 10}\, \frac{1}{U}\left\langle
\int_{0}^{+\infty}\int_{0}^{\pi/2}
\delta\left\lbrack\log_{10}U-(\cos\theta)^{1/\alpha(\varphi ,\hat{\bm{s}}_{\perp})} (a\eta -b)\right\rbrack f(\theta)h(\eta)\, d\theta d\eta
\right\rangle_{\varphi ,\hat{\bm{s}}_{\perp}} \nonumber \\
&=&\frac{1}{\ln 10}\, \frac{1}{U}\left\langle
\int_{0}^{\pi/2}\frac{f(\theta)}{a\, (\cos\theta)^{1/\alpha(\varphi ,\hat{\bm{s}}_{\perp})}}
\, h\left(\frac{\log_{10}U}{a\, (\cos\theta)^{1/\alpha(\varphi ,\hat{\bm{s}}_{\perp})}} +\frac{b}{a}\right)\, d\theta
\right\rangle_{\varphi ,\hat{\bm{s}}_{\perp}} \nonumber \\
&=&\left\langle
\int_{0}^{\pi/2}p(U,\theta\vert\varphi ,\hat{\bm{s}}_{\perp})\, d\theta
\right\rangle_{\varphi ,\hat{\bm{s}}_{\perp}},
\end{eqnarray}
with
\begin{equation}\label{weightrho}
p(U,\theta\vert\varphi ,\hat{\bm{s}}_{\perp})=\frac{1}{\ln 10}\, 
\frac{f(\theta)}{Ua\, (\cos\theta)^{1/\alpha(\varphi ,\hat{\bm{s}}_{\perp})}}
\, h\left(\frac{\log_{10}U}{a\, (\cos\theta)^{1/\alpha(\varphi ,\hat{\bm{s}}_{\perp})}} +\frac{b}{a}\right).
\end{equation}
The expression for $p(\mathcal{A}, U)$ is similar to equation~(\ref{pofuintegral}) with integrand $\bm{1}_{\bm{s}\in\mathcal{A}}\, p(U,\theta\vert\varphi ,\hat{\bm{s}}_{\perp})$, where $\bm{s}=\sqrt{\eta} \cos\theta\, {\rm e}^{i\varphi} \bm{\mathfrak{e}}^{(1)} + \sqrt{\eta} \sin\theta\, \hat{\bm{s}}_{\perp}$ with $\eta=a^{-1}(b+\log_{10}U/(\cos\theta)^{1/\alpha(\varphi,\hat{\bm{s}}_{\perp})})$. Dividing this expression by $p(U)$ one obtains
\begin{equation}\label{conditionalproba}
P(\mathcal{A}\vert U)=
\left\langle w_{\alpha(\varphi ,\hat{\bm{s}}_{\perp})}(U)
\int_{0}^{\pi/2} \bm{1}_{\bm{s}\in\mathcal{A}}\, p(\theta\vert U,\varphi ,\hat{\bm{s}}_{\perp})\, d\theta
\right\rangle_{\varphi ,\hat{\bm{s}}_{\perp}},
\end{equation}
with
\begin{equation}\label{interdef1}
p(\theta\vert U,\varphi ,\hat{\bm{s}}_{\perp})=
\frac{p(U,\theta\vert\varphi ,\hat{\bm{s}}_{\perp})}{p(U\vert\varphi ,\hat{\bm{s}}_{\perp})},
\end{equation}
and
\begin{equation}\label{interdef2}
w_{\alpha(\varphi ,\hat{\bm{s}}_{\perp})}(U)=
\frac{p(U\vert\varphi ,\hat{\bm{s}}_{\perp})}{p(U)}.
\end{equation}
 In equations~(\ref{interdef1}) and (\ref{interdef2}), $p(U\vert\varphi ,\hat{\bm{s}}_{\perp})$ is obtained by integrating equation~(\ref{weightrho}) numerically over $0\le\theta\le\pi/2$ (for fixed $U$ and a given realization of $\varphi$ and $\hat{\bm{s}}_{\perp}$). Equations~(\ref{pofuintegral}) to (\ref{interdef2}) give the expressions of $p(U)$ and $P(\mathcal{A}\vert U)$ valid for $\log_{10}U\gg 1$.
%
%
\section{Biased sampling and numerical validation of instanton analysis}\label{application}
\begin{figure}[!h]
\includegraphics[width = 0.6\linewidth]{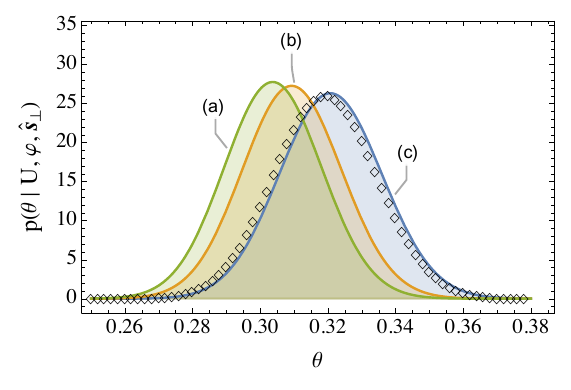}
\caption{Plots of $p(\theta\vert U,\varphi ,\hat{\bm{s}}_{\perp})$ for fixed $\log_{10}U=1890$ and three particular realizations in $\lbrace\varphi ,\hat{\bm{s}}_{\perp}\rbrace$ corresponding to the smallest (a), middle (b), and largest (c) values of $p(U\vert\varphi ,\hat{\bm{s}}_{\perp})$. Plot of $p(\theta\vert U)$ for the same value of $\log_{10}U=1890$ (diamonds).}
\label{figurerho}
\end{figure}

Let $\lbrace\varphi ,\hat{\bm{s}}_{\perp}\rbrace$ denote the same sample of $10^2$ independent realizations of $\varphi$ and $\hat{\bm{s}}_{\perp}$ as the one we used to check the validity of the nonlinear fit in equation~(\ref{nonlinearfit}). Figure\ \ref{figurerho} shows plots of $p(\theta\vert U,\varphi ,\hat{\bm{s}}_{\perp})$ for three different realizations in $\lbrace\varphi ,\hat{\bm{s}}_{\perp}\rbrace$ and $\log_{10}U=1890$. Curves (a), (b), and (c) correspond to the realizations of $\lbrace\varphi ,\hat{\bm{s}}_{\perp}\rbrace$ yielding the smallest, middle, and largest values of $p(U\vert\varphi ,\hat{\bm{s}}_{\perp})$, respectively. Diamonds correspond to $p(\theta\vert U)$ computed as the sample mean of $p(\theta\vert U,\varphi ,\hat{\bm{s}}_{\perp})$ for the realizations in $\lbrace\varphi ,\hat{\bm{s}}_{\perp}\rbrace$.

For a given $\log_{10}U\gg 1$ and realizations of $\varphi$ and $\hat{\bm{s}}_{\perp}$ in $\lbrace\varphi ,\hat{\bm{s}}_{\perp}\rbrace$, the statistically significant realizations of $S$ are those for which $\theta$ is in the bulk of $p(\theta\vert U, \varphi ,\hat{\bm{s}}_{\perp})$. Figure~\ref{figurerho} and similar results obtained for different values of $U$ with $1820\le\log_{10}U\le 1940$ show that a good sampling of the corresponding significant values of $\theta$, requires probing the whole range $0.25\le\theta\le 0.38$. Since the equation~(\ref{pdftheta}) yields $P(0.25\le\theta\le 0.38)=5.2\ 10^{-86}$, it is clear that having $\theta$ in that range is an extremely rare event, virtually impossible to sample directly by drawing $\theta$ from $f(\theta)$. By contrast, it can easily be achieved by drawing $\theta$ from $p(\theta\vert U,\varphi ,\hat{\bm{s}}_{\perp})$ rather than from $f(\theta)$. This is what defines our biased sampling procedure which consists of the following four steps:

\begin{itemize}
\item[(A)]{draw $\varphi$ and $\hat{\bm{s}}_{\perp}$ from the uniform distributions over $\lbrack 0,2\pi)$ and the sphere $\|\bm{s}_{\perp}\| =1$, respectively;}
\item[(B)]{compute $\alpha(\varphi ,\hat{\bm{s}}_{\perp})$ by making the graph of the function in equation~(\ref{nonlinearfit}) fit the data in the $(\log_{10}U,\theta)$ region of interest, for some fixed $\eta$. (We have checked that for $1820\le\log_{10}U\le 2030$, $\eta$ can be chosen arbitrarily between $1010$ and $1100$;)}
\item[(C)]{use the result in equation~(\ref{weightrho}) to get $p(U,\theta\vert\varphi ,\hat{\bm{s}}_{\perp})$. For fixed $U$ (with $\log_{10}U\gg 1$), integrate the result numerically over $0\le\theta\le\pi/2$ to get $p(U\vert\varphi ,\hat{\bm{s}}_{\perp})$. Then, draw $\theta$ from $p(\theta\vert U,\varphi ,\hat{\bm{s}}_{\perp})$ in equation~(\ref{interdef1}) and set $\eta$ by inverting equation~(\ref{nonlinearfit});}
\item[(D)]{the outcome defines a realization of $\bm{s}=\sqrt{\eta} \cos\theta\, {\rm e}^{i\varphi} \bm{\mathfrak{e}}^{(1)} + \sqrt{\eta} \sin\theta\, \hat{\bm{s}}_{\perp}$ which, once injected onto the right-hand side of equation~(\ref{solSnum}), gives a realization of $S(x,z)$.}
\end{itemize}

\begin{figure}[!h]
\includegraphics[width = 0.6\linewidth]{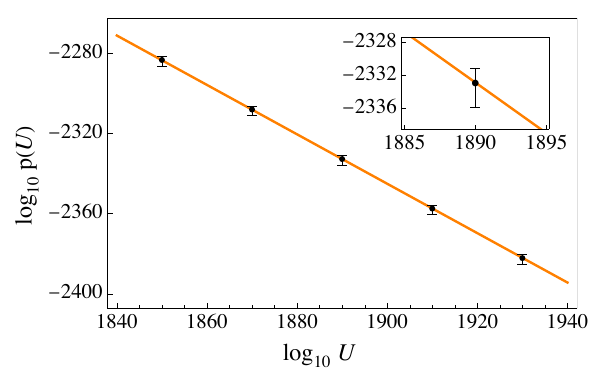}
\caption{$\log_{10}p(U)$ a a function of $\log_{10}U$ from $1850$ to $1930$ by steps of $20$, with $p(U)$ given by equation~(\ref{pofuestimated}) (black dots). Smallest and largest $p(U\vert\varphi ,\hat{\bm{s}}_{\perp})$ in the sample are indicated by the ends of vertical bars. Plot of $\log_{10}p(U)=-\zeta\log_{10}U +{\rm Const}$ with $\zeta=(1+1/2\mu_{\rm max} g)=1.22989$ given by instanton analysis and ${\rm Const}=-8.41526$ adjusted to get the best fit to numerical data (solid line). Inset: enlargement of the same plot near $\log_{10}U=1890$.}
\label{figurepofu}
\end{figure}
From the data for $p(U\vert\varphi ,\hat{\bm{s}}_{\perp})$ obtained as explained in step (C) for each realization in $\lbrace\varphi ,\hat{\bm{s}}_{\perp}\rbrace$ and fixed $U$, we have estimated $p(U)$ as the sample mean
\begin{equation}\label{pofuestimated}
p(U)=\frac{1}{{\rm card}\lbrace\varphi ,\hat{\bm{s}}_{\perp}\rbrace}
\, \sum_{(\varphi ,\hat{\bm{s}}_{\perp})\in\lbrace\varphi ,\hat{\bm{s}}_{\perp}\rbrace}
p(U\vert\varphi ,\hat{\bm{s}}_{\perp}).
\end{equation}
Figure\ \ref{figurepofu} shows the results in the $(\log_{10}U,\log_{10}p(U))$ plane for five different values of $U$ with $\log_{10}U$ between $1840$ and $1940$. Black dots correspond to $p(U)$. The dispersion of $p(U\vert\varphi ,\hat{\bm{s}}_{\perp})$ around $p(U)$ is indicated by vertical bars the ends of which correspond to the smallest and largest values of $p(U\vert\varphi ,\hat{\bm{s}}_{\perp})$ in the sample. The error bars corresponding to the standard deviation of the sample mean~(\ref{pofuestimated}) are found to be eight times shorter, in the case of figure~\ref{figurepofu} (not shown). Instanton analysis predicts a leading algebraic tail of $p(U)\propto U^{-\zeta}$ with $\zeta=(1+1/2\mu_{\rm max} g)=1.22989$. The solid line is the plot of $\log_{10}p(U)=-\zeta\log_{10}U +{\rm Const}$, where ${\rm Const}=-8.41526$ has been adjusted to get the best fit to the data. We observe an almost perfect alignment of numerical data along a straight line with slope $-\zeta$, which validates the result of instanton analysis numerically in the considered range of $\log_{10}U$. Note also the extremely small value of $p(U)<10^{-2270}$ which confirms, if need be, the absolute impossibility of sampling the extreme upper tail of $p(U)$ directly, without a specific bias procedure.

\begin{figure}[!h]
\includegraphics[width = 0.6\linewidth]{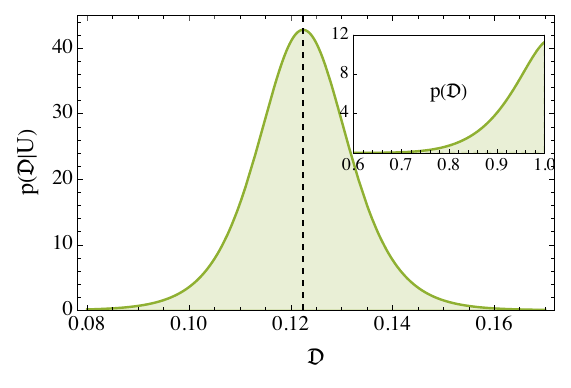}
\caption{$p(\mathfrak{D}\vert U)$ for fixed $\log_{10}U=1890$ estimated from a biased sample of $10^6$ realizations of $S$ drawn according to the biased sampling procedure defined in steps (A) to (D) (see the text for details). The median of $p(\mathfrak{D}\vert U)$ is at $\mathfrak{D}=0.122457$ (dashed line). Inset: $p(\mathfrak{D})$ estimated from an unbiased sample of $10^5$ realizations of $S$.}
\label{figurepofdl2}
\end{figure}
The question then arises of the realizations of $S$ behind the results in figure\ \ref{figurepofu}. According to the instanton analysis in\ \cite{Mounaix2023}, these realizations should be instanton realizations with the same profile as in figure\ \ref{figureinst}(a). For a given $U$, the way $S$ differs from the instanton can be characterized by the conditional pdf $p(\mathfrak{D}\vert U)$, where $\mathfrak{D}$ is the $L^2$-distance defined in equation~(\ref{distnum2}). For each element of $\lbrace\varphi ,\hat{\bm{s}}_{\perp}\rbrace$ and fixed $U$, we drew $10^4$ independent realizations of $\theta$ --- denoted in the following by $\lbrace\theta\vert\varphi ,\hat{\bm{s}}_{\perp}\rbrace$ --- as explained in step (C). For definiteness, we took $\log_{10}U=1890$ at the center of the range considered in figure\ \ref{figurepofu}. There are $10^2$ subsamples $\lbrace\theta\vert\varphi ,\hat{\bm{s}}_{\perp}\rbrace$ of $10^4$ elements each, representing a total of $10^6$ different realizations of $\bm{s}$. For each $\bm{s}$, we have used the equation~(\ref{distnum2}) to compute the corresponding value of $\mathfrak{D}$. We have then estimated $P(\mathfrak{D}\le\delta\vert U)$ in equation~(\ref{conditionalproba}) with $\mathcal{A}=\lbrace\bm{s}\vert\mathfrak{D}\le\delta\rbrace$ as the sample mean
\begin{equation}\label{pofdlessdelta}
P(\mathfrak{D}\le\delta\vert U)=\frac{1}{{\rm card}\lbrace\varphi ,\hat{\bm{s}}_{\perp}\rbrace}
\, \sum_{(\varphi ,\hat{\bm{s}}_{\perp})\in\lbrace\varphi ,\hat{\bm{s}}_{\perp}\rbrace}
\, \frac{w_{\alpha(\varphi ,\hat{\bm{s}}_{\perp})}(U)}{{\rm card}\lbrace\theta\vert\varphi ,\hat{\bm{s}}_{\perp}\rbrace}
\, \sum_{\theta\in\lbrace\theta\vert\varphi ,\hat{\bm{s}}_{\perp}\rbrace}
\bm{1}_{\mathfrak{D}\le\delta},
\end{equation}
from which $p(\mathfrak{D}\vert U)$ is obtained by (numerical) derivation with respect to $\delta$ at $\delta=\mathfrak{D}$. Figure~\ref{figurepofdl2} shows the result for $\log_{10}U=1890$. For comparison, we show in inset the pdf of $\mathfrak{D}$ obtained from an unbiased sample of $10^5$ realizations of $S$. The vertical dashed line indicates the median of $p(\mathfrak{D}\vert U)$ at $\mathfrak{D}=\tilde{\mathfrak{D}}=0.122457$. (The median, mean and maximum of $p(\mathfrak{D}\vert U)$ are all at $\tilde{\mathfrak{D}}$, to within numerical accuracy.) It can be seen that the realizations of $S$ conditioned to a large $\log_{10}U\gg 1$ are significantly closer to the instanton than unconditioned realizations: $0.08<\mathfrak{D}<0.16$ and $0.6<\mathfrak{D}\le 1$, respectively, in the case of figure~\ref{figurepofdl2}. This is in agreement with the instanton analysis in \cite{Mounaix2023} which predicts $\mathfrak{D}\rightarrow 0$ in probability, as $\log_{10}U\rightarrow +\infty$.

\begin{figure}[!h]
\includegraphics[width = 0.9\linewidth]{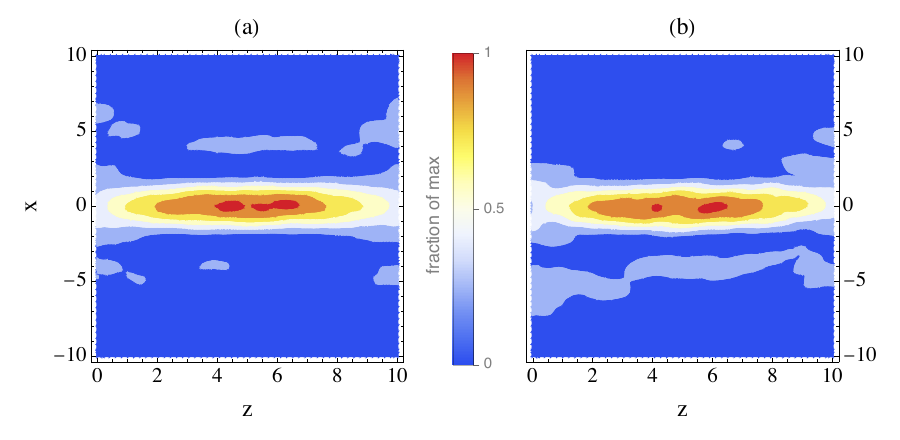}
\caption{Contour plots of $\vert\hat{S}(x,z)\vert^2$ for two realizations of $S$ with $\mathfrak{D}\simeq 0.08$ (a) and $\mathfrak{D}\simeq 0.16$ (b).}
\label{figuresprofileminmax}
\end{figure}
\begin{figure}[!h]
\includegraphics[width = 0.9\linewidth]{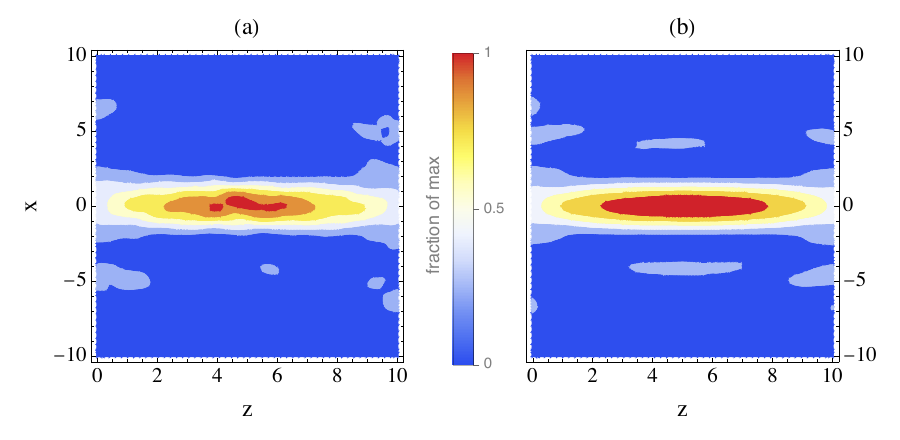}
\caption{(a) Contour plot of $\vert\hat{S}(x,z)\vert^2$ for a realization of $S$ with $\mathfrak{D}\simeq\tilde{\mathfrak{D}}$. (b) Contour plot of the sample mean of $\vert\hat{S}(x,z)\vert^2$ for the $10^2$ realizations in $\lbrace S\rbrace_{\mathfrak{D}=\tilde{\mathfrak{D}}}$ (with $0.122403\le\mathfrak{D}\le 0.122504$).}
\label{figuresamplemean}
\end{figure}

In figure\ \ref{figuresprofileminmax}, we show $\vert\hat{S}(x,z)\vert^2=\vert S(x,z)\vert^2/\| S\|_2^2$ for two realizations of $S$ with $\mathfrak{D}\simeq 0.08$ and $\mathfrak{D}\simeq 0.16$, on both sides of the bulk of $p(\mathfrak{D}\vert U)$. Figure~\ref{figuresamplemean}(a) shows the same quantity for a realization with $\mathfrak{D}\simeq\tilde{\mathfrak{D}}$, right at the center of the bulk of $p(\mathfrak{D}\vert U)$. The results in figures~\ref{figuresprofileminmax}(a), \ref{figuresprofileminmax}(b), and \ref{figuresamplemean}(a) are very similar and typical of the realizations generated by the biased sampling procedure for $\log_{10}U=1890$. These realizations are the superposition of an elongated cigar-shaped component along $x=0$ and fluctuations of comparatively small amplitude. Fluctuations can be smoothed out by averaging realizations of $\vert\hat{S}\vert^2$ close together along the $\mathfrak{D}$ axis, bringing out the underlying cigar-shaped component. To this end, we have constructed five subsamples $\lbrace S\rbrace_{\mathfrak{D}=\delta}$ of $10^2$ realizations of $S$ selected by picking in the total (biased) sample the $50$ realizations with largest values of $\mathfrak{D}\le\delta$ and the $50$ realizations with smallest values of $\mathfrak{D}>\delta$, for $\delta =0.1$, $0.11$, $\tilde{\mathfrak{D}}\, (=0.122457)$, $0.13$, and $0.14$. Then, we have computed the sample means of $\vert\hat{S}(x,z)\vert^2$ for the realizations in each $\lbrace S\rbrace_{\mathfrak{D}=\delta}$. In figure\ \ref{figuresamplemean}(b), we show the result for $\delta =\tilde{\mathfrak{D}}$. (The smallest and largest values of $\mathfrak{D}$ for $S$ in $\lbrace S\rbrace_{\mathfrak{D}=\tilde{\mathfrak{D}}}$ are $0.122403$ and $0.122504$, respectively.) We have obtained identical results for the five different values of $\delta$ we have considered.

The close resemblance between figures~\ref{figuresamplemean}(b) and \ref{figureinst}(a) is obvious and the dominant cigar-shaped component of $\vert\hat{S}\vert^2$ along $x=0$ visible in figures~\ref{figuresprofileminmax}(a), \ref{figuresprofileminmax}(b), and \ref{figuresamplemean}(a) can clearly be identified as the theoretical instanton profile. These results, together with the small values of $\mathfrak{D}$ observed in figure~\ref{figurepofdl2}, show that the realizations of $S$ conditioned on a large but finite $\log_{10}U$ (here, $\log_{10}U=1890$) are low-noise instanton realizations. This is in agreement with the instanton analysis in \cite{Mounaix2023} which predicts noiseless instanton realizations in the limit $\log_{10}U\rightarrow +\infty$.

The numerical results in figures~\ref{figurepofu} to \ref{figuresamplemean} show remarkable agreement with the analytical predictions of instanton analysis. Data points for $\log_{10}p(U)$ and $\log_{10}U$ line up almost perfectly along the predicted algebraic tail $\log_{10}p(U)\simeq -\zeta\log_{10}U$ with $\zeta=(1+1/2\mu_{\rm max} g)$ (figure~\ref{figurepofu}), and the corresponding realizations of $S$ are observed to be heavily biased towards the predicted instanton realizations (figures~\ref{figurepofdl2} to \ref{figuresamplemean}). In conclusion, we can say that our results provide a convincing numerical validation of the instanton analysis in the large amplification regime $\log_{10}U\gg 1$.
%
%
\section{Summary and discussion}\label{summary}
In this paper, we have numerically tested the validity of the instanton analysis approach to study the diffraction-amplification problem~(\ref{withDeq}) in the large amplification regime $\ln U\gg 1$, where $U=\vert\psi(0,L)\vert^2$, $\psi(x,z)$ being the solution to equation~(\ref{withDeq}). By analyzing a large number of numerical solutions to equation~(\ref{withDeq}) with $S$ in the `one-max class' defined at the beginning of Section~\ref{subsec3a}, we have identified a nonlinear fit to numerical data from which a highly accurate approximation of $U$ as a function of $S$ can be obtained, when $\ln U$ is large (equation~(\ref{nonlinearfit})). We have then used this result to devise a sampling procedure of $S$ giving access to large values of $\ln U$.

As a first application, we have obtained $p(U)$ numerically over a large range of $U$ with $\ln U\gg 1$, down to probability density less than $10^{-2270}$ in the tail. We have found a near-perfect agreement with the algebraic tail of $p(U)$ theoretically predicted by the instanton analysis in\ \cite{Mounaix2023} (figure~\ref{figurepofu}). Then, we have determined the conditional probability distribution of $\mathfrak{D}$ given $U$ for a large $\ln U$, where $\mathfrak{D}$ is the $L^2$-distance measuring the departure of $S$ from the theoretical instanton normalized such that $0\le\mathfrak{D}\le 1$. We have found that the realizations of $S$ in the far right tail of $p(U)$ are significantly closer to the instanton than unconditioned realizations: $0.08<\mathfrak{D}<0.16$ and $0.6<\mathfrak{D}\le 1$ respectively, in the case of figure~\ref{figurepofdl2}. As a confirmation of this result, plots of $\vert S\vert^2/\| S\|^2_{2}$ clearly show that the realizations of $S$ in the far right tail of $p(U)$ are low-noise instanton realizations around the predicted, noiseless, theoretical instanton, with residual noise due to $\ln U$ being finite (figures~\ref{figuresprofileminmax} and \ref{figuresamplemean}). To summarize, our numerical results validate the instanton analysis of the diffraction-amplification problem~(\ref{withDeq}) in the large $\ln U$ limit for $S$ in the one-max class.

In conclusion we briefly discuss the possibility of using more standard sampling methods to deal with the same problem. What makes our approach particularly efficient (compared with standard ones) is that we do not have to know the maximizing direction prior to testing whether any given direction is close to it or not. All we need is to look at how $\ln U$ depends on ${\bf s}$ around the tested direction (here, the instanton direction). This saves a considerable amount of time as we do not have to set off in search of where the maximizing direction is on the $2N$-dimensional unit sphere (with $N=100$ for $S$ in equation~(\ref{solSnum})). Either the observed behavior of $\ln U$ indicates that the instanton direction is close to the maximizing direction, and we can move on to the biased sampling procedure around the instanton direction, as explained in Section\ \ref{application}, or it does not and the instanton solution must be rejected. By contrast, a standard algorithm like in, e.g.,\ \cite{Hartmann2014,HDMRS2018,HMS2019} involves probing the $2N$-dimensional unit sphere until it converges to the maximizing direction, which requires massive numerical simulations. For each $\hat{\bm{s}}$ probed, problem (\ref{withDeq}) with the corresponding $S$ must be solved numerically, and a great many $\hat{\bm{s}}$'s are likely to be probed before converging to the maximizing direction. (Unless prior knowledge of the instanton to be tested is used to pick a starting direction not too far from the maximizing one.)

On the other hand, more standard approaches can in principle give access to large values of $\ln U$ in cases where no analytical prediction is available, unlike the method presented here. They may therefore be used when instanton analysis is too difficult (if not impossible) to carry out. In such cases, the landscape of $\ln U(\eta,\hat{\bm s})$ to be explored is a priori totally unknown, which can pose tricky problems the solution of which is likely to make the algorithm even more numerically demanding. In this respect, other general sampling algorithms like, e.g., subset simulation\ \cite{AB2001,PBZS2015}, may also be of interest as possible alternatives to important sampling algorithms.

To the best of the author's knowledge, general iterative algorithms for sampling rare event sets have never been used so far to deal with stochastic amplifiers in the large amplification limit. Unless there are hidden fundamental reasons preventing such use, it would open up a very interesting new field to explore. This will be the subject of a future work.
%
%
\acknowledgments{The author warmly thanks Denis Pesme for his interest and valuable advice about the manuscript. He also thanks the anonymous referee of reference\ \cite{Mounaix2023} whose pertinent and constructive remarks have motivated this work.}
%
%
\appendix
\section{A sufficient condition for $\bm{S}$ to belong to the one-max class}\label{app1}
The class of $S$ considered in the paper is defined by the condition that, in the large $\eta$ limit, there is only one direction of $\bm{s}$ that globally maximizes $\ln U(\eta ,\hat{\bm{s}})$. In this appendix, we give a sufficient condition ensuring that a given $S$ do belong to this class.

For any spectral density $\varsigma_n$ (not necessarily Gaussian), we take
\begin{equation}\label{KLexpansion}
S(x,z)=\sum_{\mathclap{n=-N/2}}^{N/2}{\vphantom{\sum}}^\prime
s_{n}\sqrt{\varsigma_n}\, 
{\rm e}^{2i\pi nx/\ell}\Phi_n(z),
\end{equation}
which generalizes the equation~(\ref{solSnum}), where $N$ is an even integer and the $\Phi_n$'s are continuous function of $0\le z\le L$. Let $B(0,L)$ denote the set of all the continuous paths in $\Lambda$ satisfying $x(L)=0$ and define $M\lbrack x(\cdot)\rbrack$ the $N\times N$ positive definite matrix with components
\begin{equation}\label{generalmatrixM}
M_{nm}\lbrack x(\cdot)\rbrack =\sqrt{\varsigma_n \varsigma_m}
\int_0^L{\rm e}^{2i\pi(m-n)x(z)/\ell}\Phi_n(z)^\ast\Phi_m(z)\, dz,
\end{equation}
in which $x(\cdot)\in B(0,L)$. Writing the solution to the equation~(\ref{withDeq}) formally as a Feynman-Kac integral over $B(0,L)$ and using the equations~(\ref{KLexpansion}) and (\ref{generalmatrixM}), one gets
\begin{eqnarray}\label{FKsolution}
\psi(0,L)&=&\int_{x(\cdot)\in B(0,L)}{\rm e}^{\int_{0}^{L}
\left\lbrack\frac{im}{2}\dot{x}(\tau)^2+g\vert S(x(\tau),\tau)\vert^2\right\rbrack\, d\tau}
\mathscr{D}x \nonumber \\
&=&\int_{x(\cdot)\in B(0,L)}{\rm e}^{\int_{0}^{L}
\left\lbrack\frac{im}{2}\dot{x}(\tau)^2+g\eta\, \hat{\bm{s}}^\dagger M\lbrack x(\cdot)\rbrack\hat{\bm{s}}\right\rbrack\, d\tau}
\mathscr{D}x.
\end{eqnarray}
In the large $\eta$ limit, the amplification of $\psi$ is dominated by the most amplified paths, and one has
\begin{equation}\label{modulepsi2}
\vert\psi(0,L)\vert^2\asymp\exp\left\lbrack 2g\eta\, 
\sup_{x(\cdot)\in B(0,L)}\hat{\bm{s}}^\dagger M\lbrack x(\cdot)\rbrack\hat{\bm{s}}\right\rbrack ,
\end{equation}
where the symbol $\asymp$ means asymptotic logarithmic equivalence; i.e., the ratio of the logarithms of the two sides tends to $1$ as $\eta\to +\infty$. Writing $\mu_1\lbrack x(\cdot)\rbrack$ the largest eigenvalue of $M\lbrack x(\cdot)\rbrack$, it follows immediately from the equation~(\ref{modulepsi2}) that
\begin{eqnarray}\label{maxofU}
\max_{\hat{\bm{s}}}\ln U(\eta ,\hat{\bm{s}})&=&
\max_{\hat{\bm{s}}}\ln\vert\psi(0,L)\vert^2 \nonumber \\
&\sim&2g\eta\, \max_{\hat{\bm{s}}}
\sup_{x(\cdot)\in B(0,L)}\hat{\bm{s}}^\dagger M\lbrack x(\cdot)\rbrack\hat{\bm{s}} \\
&=&2g\eta\, \sup_{x(\cdot)\in B(0,L)}\mu_1\lbrack x(\cdot)\rbrack\ \ \ \ \ (\eta\to +\infty), \nonumber
\end{eqnarray}
the maximum being reached at $\hat{\bm{s}}$ along the fundamental eigenvectors of $M\lbrack x_{\rm inst}(\cdot)\rbrack$, where $x_{\rm inst}$ is a path of $B(0,L)$ maximizing $\mu_1\lbrack x(\cdot)\rbrack$. We recall that for the $S$'s considered in\ \cite{Mounaix2023}, all the paths maximizing $\mu_1\lbrack x(\cdot)\rbrack$ are in $B(0,L)$ and there is a finite number of such paths. The interested reader will find a mathematically rigorous demonstration of (\ref{maxofU}) in\ \cite{MCL2006}.

If none of the $\mu_1\lbrack x_{\rm inst}(\cdot)\rbrack$'s is degenerate and all the corresponding normalized eigenvectors are equal, then $S$ admits a single and non-degenerate instanton (by definition). In this case, there is only one direction $\hat{\bm{s}}$ at which the maximum of $\ln U(\eta ,\hat{\bm{s}})$ in (\ref{maxofU}) is reached. Namely, the direction of the fundamental eigenvector common to all the $\mu_1\lbrack x_{\rm inst}(\cdot)\rbrack$'s. Therefore, admitting a single and non-degenerate instanton is a sufficient condition for $S$ to belong to the one-max class. It is shown in\ \cite{Mounaix2023} that for $S$ given by the equations~(\ref{solSnum}) and (\ref{gaussspectrum}), there is only one path, $x_{\rm inst}(\cdot)\equiv 0$, that maximizes $\mu_1\lbrack x(\cdot)\rbrack$, and the corresponding eigenvalue $\mu_1\lbrack x_{\rm inst}(\cdot)\equiv 0\rbrack$ is non-degenerate. Thus, $S$ in equations~(\ref{solSnum}) and (\ref{gaussspectrum}) admits a single and non-degenerate instanton, hence it belongs to the one-max class.
%
%

%
\end{document}